\documentstyle[aps,prb]{revtex}

\input{epsf}
\newcommand{\bge}{\begin{equation}}
\newcommand{\ee}{\end{equation}}
\newcommand{\bgea}{\begin{eqnarray}}
\newcommand{\eea}{\end{eqnarray}}
\newcommand{\bgeas}{\begin{eqnarray*}}
\newcommand{\eeas}{\end{eqnarray*}}
\newcommand{\bfci}{{\bf c}_i}
\newcommand{\bfcj}{{\bf c}_j}
\newcommand{\bfu}{{\bf u}}
\newcommand{\bfx}{{\bf x}}
\newcommand{\mixten}[3]{{#1}^{#2}_{\phantom{{#2}} #3}}
\newcommand{\smallhalf}{\mbox{\small $\frac{1}{2}$}}
\def\bfalpha{\mbox{\boldmath $\alpha$}}

\def\bfnabla{\mbox{\boldmath $\nabla$}}

\begin{document}

\title{
\begin{flushleft}
{\small BU-CCS-960101}\\
\end{flushleft}
Integer Lattice Gases
}
\author{
Bruce M. Boghosian\\
{\small \sl Center for Computational Science,} \\
{\small \sl Boston University,} \\
{\small \sl 3 Cummington Street, Boston, Massachusetts 02215, U.S.A.} \\
{\small \tt bruceb@bu.edu} \\[0.3cm]
Jeffrey Yepez\\
{\small \sl Phillips Laboratory,} \\
{\small \sl Hanscom AFB, Massachusetts, U.S.A.} \\
{\small \tt yepez@wave.plh.af.mil} \\[0.3cm]
Francis J. Alexander\\
{\small \sl Center for Computational Science,} \\
{\small \sl Boston University,} \\
{\small \sl 3 Cummington Street, Boston, Massachusetts 02215, U.S.A.} \\
{\small \tt fja@bu.edu} \\[0.3cm]
Norman H. Margolus\\
{\small \sl Laboratory for Computer Science,} \\
{\small \sl Massachusetts Institute of Technology,} \\
{\small \sl Technology Square, Cambridge, Massachusetts 02139, U.S.A.} \\
{\small \tt nhm@im.lcs.mit.edu} \\[0.3cm]
}
\date{\today}
\maketitle

\begin{abstract}
  We generalize the hydrodynamic lattice gas model to include arbitrary
  numbers of particles moving in each lattice direction.  For this
  generalization we derive the equilibrium distribution function and the
  hydrodynamic equations, including the equation of state and the
  prefactor of the inertial term that arises from the breaking of
  galilean invariance in these models.  We show that this prefactor can
  be set to unity in the generalized model, therby effectively restoring
  galilean invariance.  Moreover, we derive an expression for the
  kinematic viscosity, and show that it tends to decrease with the
  maximum number of particles allowed in each direction, so that higher
  Reynolds numbers may be achieved.  Finally, we derive expressions for
  the statistical noise and the Boltzmann entropy of these models.
\end{abstract}

\section{Lattice Gases}

Lattice gas automata (LGA) are a class of dynamical systems in which
particles move on a lattice in discrete time steps.  If the collisions
between the particles conserve mass and momentum, the coarse-grained
behavior of the system can be shown to be that of a viscous fluid in the
appropriate scaling limit~\cite{bib:fhp,bib:swolf,bib:fchc,bib:long}.
Used as an algorithm for simulating hydrodynamics, the method has the
virtues of {\it exact} conservation laws, and of {\it unconditional}
numerical stability.

In a typical LGA, there is an association between the lattice vectors
and the particles at each site.  If there are $n$ lattice vectors, then
the state of the site is represented by $n$ bits.  Each bit represents
the presence or absence of a particle in the corresponding direction.
At each time step, a particle propagates along its corresponding lattice
vector and then collides with other arriving particles at the new
site\footnote{Note that rest particles can be subsumed into this scheme
by associating them with null lattice vectors.}.  The collisions are
required to conserve particle mass and momentum.

Relevant dimensionless quantities of a LGA are the Knudsen number,
$\mbox{Kn}$, defined as the ratio of the mean-free path to the
characteristic length scale; the Strouhal number, $\mbox{Sh}$, defined
as the ratio of the mean-free time to the characteristic time scale; the
Mach number, $\mbox{M}$, defined as the ratio of the characteristic
velocity to the speed of sound; the Reynolds number,
$\mbox{Re}\sim\mbox{M}/\mbox{Kn}$; and the fractional variation of
density from its average value, $\delta\rho/\rho$.  Hydrodynamic
behavior~\cite{bib:cerc} is attained in the limit as $\mbox{Kn}$ and
$\mbox{Sh}$ go to zero.  {\it Viscous} hydrodynamics~\cite{bib:cerc} is
attained when $\mbox{Sh}\sim\mbox{Kn}^2$ in this limit.  {\it
Incompressible} viscous hydrodynamics~\cite{bib:ll} is then attained
when we also have $\mbox{M}\sim\mbox{Kn}$ so that $\mbox{Re}\sim {\cal
O}(1)$, and $\delta \rho/\rho\sim\mbox{Kn}^2$.

The Chapman-Enskog procedure is a perturbation expansion in the
above-described asymptotic ordering.  For a LGA whose collisions
conserve mass and momentum on a lattice of sufficient symmetry
(quantified below), the local equilibrium distribution function can be
shown to be Fermi-Dirac in nature~\cite{bib:swolf,bib:fchc,bib:long}.
The Chapman-Enskog procedure can then be used to compute the correction
to this Fermi-Dirac distribution and thereby
show~\cite{bib:fhp,bib:swolf,bib:fchc,bib:long} that the pressure, $P$,
and the momentum density, $\bfu$, obey the following equations in the
asymptotic limit:
\[
\bfnabla\cdot\bfu = 0
\]
\[
\frac{\partial\bfu}{\partial t} +
 \frac{g(\rho)}{\rho}\bfu\cdot\bfnabla\bfu =
 -\bfnabla P + \nu (\rho)\nabla^2\bfu,
\]
where $\rho$ is the fluid density (a constant in this limit).  The
analysis also yields expressions for the functions $g(\rho)$ and
$\nu(\rho)$, and an equation of state for $P$.  In particular, the form
of these equations, the equation of state, and the expression for the
function $g(\rho)$ depend only on the fact that mass and momentum are
conserved -- and are the only things conserved -- by the collisions.
The expression for the viscosity, $\nu(\rho)$ depends on the details of
the collision rules used.

Since the fluid density $\rho$ is a constant in the asymptotic limit,
the factors $g(\rho)$ and $\nu (\rho)$ are also constants.  As has been
noted, the latter is the fluid viscosity.  The presence of the former is
reflective of a breaking of galilean invariance, due to the fact that
the lattice itself constitutes a preferred galilean frame of reference.
For a single-phase LGA, the former factor can easily be scaled away by
redefining the momentum density and pressure as
\[
{\bf U}\equiv g(\rho)\bfu
\]
and
\[
{\cal P}\equiv g(\rho) P,
\]
where $\bfu$ and $P$ are those measured in the simulation.  For
compressible flow, or for multiphase flow with interfaces, however, the
presence of the $g(\rho)$ factor is problematic, and various techniques
have been proposed to remove it.  It has been shown that this can be
done by judiciously violating semi-detailed balance in the collision
rule~\cite{bib:dhu}, or by adding many rest particles at each
site~\cite{bib:roth}.

The unconditional stability of the lattice gas procedure arises from a
requirement that the collisions satisfy a statistical reversibility
condition known as {\it semi-detailed balance} (SDB).  The collision
process is fully specified by the transition matrix $A(s\rightarrow s')$
which is the probability that the incoming state $s$ will result in
outgoing state $s'$.  Since collisions must result in some outgoing
state, conservation of probability requires that
\begin{equation}
 \sum_{s'} A(s\rightarrow s') = 1.
 \label{eq:consprob}
\end{equation}
SDB is then the condition that
\begin{equation}
 \sum_{s} A(s\rightarrow s') = 1.
 \label{eq:sdb}
\end{equation}
(Note that the condition of {\it detailed balance} (DB), $A(s\rightarrow
s')=A(s'\rightarrow s)$, implies that of SDB, but not vice versa; that
is, SDB is a weaker condition than DB.)  From SDB, it is possible to
prove an $H$-theorem, from which follows the unconditional stability of
the lattice gas algorithm.

An important limitation of the lattice gas procedure has to do with the
statistical noise associated with the coarse-grained averaging that is
necessary to get the hydrodynamic quantities that obey the above fluid
equations.  For $n$ bits per site, and for coarse-grained averages over
blocks of $N$ sites, the noise is of order $\sim 1/\sqrt{nN}$.  For some
applications -- most notably the simulation of complex fluids -- a
certain controllable amount of noise is actually desirable because it is
essential to the physics; for simple fluid dynamics computations, on the
other hand, the noise is a nuisance.

\section{Lattice Boltzmann Equations}

Because of their noise and lack of galilean invariance, LGA have been
replaced by Lattice Boltzmann Equations (LBE) for many hydrodynamics
applications of interest in recent years~\cite{bib:lbe1}.  These methods
keep track only of an averaged occupation number of particles in each
direction at each site.  Moreover, the collision operator most often
used is a simple relaxation to a noiseless equilibrium, thereby
eliminating the statistical fluctuations that are inherent in the LGA
method.  This means that in complex fluid applications for which
statistical fluctuations are an essential part of the physics, they have
to be reintroduced artificially~\cite{bib:ladd}.

For a lattice Boltzmann equation corresponding to a lattice gas with
only one bit per lattice vector, this real-valued distribution function
is bounded between zero and one.  This need not be the case, however,
and the LBE procedure allows one to tailor the equilibrium distribution
function to satisfy certain desiderata.  Among these is the ability to
demand galilean invariance ($g(\rho)=1$)~\cite{bib:lbe2}.

At the same time, the LBE method gives up two of the principal
advantages of LGA's: Due to the roundoff error inherent in manipulations
of real numbers on a computer, it no longer maintains the conservation
laws exactly.  Moreover, LBE's are no longer unconditionally stable;
indeed, they are subject to a variety of numerical instabilities, most
of which are not well understood.

\section{Integer Lattice Gas Automata}

In this paper, we investigate a simple generalization of the lattice gas
concept that can be used to control the level of statistical
fluctuations -- reducing it if desired, but not necessarily eliminating
it altogether -- while maintaining the conservation laws exactly,
preserving unconditional stability, and allowing for galilean
invariance.

The use of a single bit per each of $n$ directions to represent the
state of a given lattice site means that the number of particles moving
along any lattice direction is either zero or one.  We generalize this
by allowing for up to $L$ bits per direction, for a total of $nL$ bits
per site, so that the number of particles moving along any lattice
direction can range from $0$ to $2^L-1$.  The total number of states per
site is then $2^{nL}$.  Computationally, this means that the state of
each direction is described by an integer of $L$ bits; hence, the
terminology, {\it Integer Lattice Gas Automata} (ILGA).

To simplify the derivation of the hydrodynamic equations of an ILGA, we
use the Boltzmann molecular chaos approximation, so that all quantities
in our analysis are ensemble averaged, and we indiscriminately commute
the application of this average with the collision process.  We also
assume that the particles are of unit mass.  Denote the
ensemble-averaged value of the $\ell$th bit in the $i$th direction by
$N^{i,\ell}$, where $0 < i < n-1$ and $0 < \ell < L-1$.  Also, denote
the lattice vector for the $i$th direction by $\bfci$.  The distribution
function for the total number of particles in each direction is then,
\bge
N^i = \sum_{\ell=0}^{L-1} 2^\ell N^{i,\ell}.
\label{eq:nidef}
\ee

The ensemble-averaged mass and momentum densities are then given by
\bge
\rho
 =
 \sum_{i=0}^{n-1} N^i
 =
 \sum_{i=0}^{n-1}\sum_{\ell=0}^{L-1} 2^\ell N^{i,\ell}
 \label{eq:masdef}
\ee
and
\bge
\bfu
 =
 \sum_{i=0}^{n-1}\bfci N^i
 =
 \sum_{i=0}^{n-1}\sum_{\ell=0}^{L-1} 2^\ell\bfci N^{i,\ell}.
 \label{eq:momdef}
\ee
Let us also associate an energy $\varepsilon_i$ with each particle in
direction $i$.  The ensemble-averaged energy density is then given by
\bge
\varepsilon
 =
 \sum_{i=0}^{n-1}\varepsilon_i N^i
 =
 \sum_{i=0}^{n-1}\sum_{\ell=0}^{L-1} 2^\ell\varepsilon_i N^{i,\ell}.
 \label{eq:enedef}
\ee

\section{Thermodynamics of the Integer Lattice Gas}
\label{sec:ent}

\begin{figure}
\epsfbox{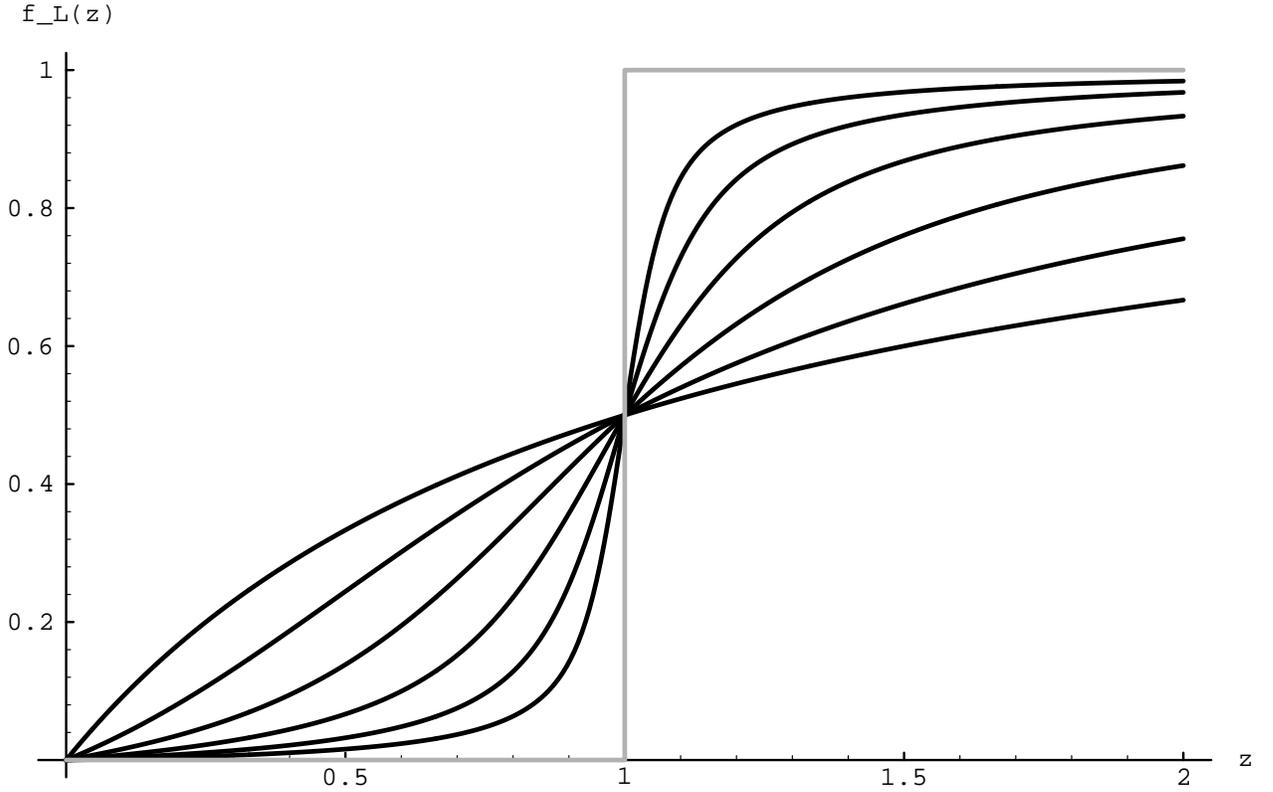}
\caption{$f_L(z)$ versus $z$ for several values of $L$.  The black
curves represent $L$ values from 1 to 6, with increasing steepness,
while the gray curve is the limit as $L\rightarrow\infty$.}
\label{fig:mop}
\end{figure}

We first consider the thermodynamics of the integer lattice gas.  The
grand canonical partition function is
\[
{\cal Z} =
 \sum_{\{N\}}
 \exp
 \left[
  -\beta
  \left(E-\bfalpha\cdot {\bf P}-\mu M\right)
 \right],
\]
where the sum is over all possible states of the ILGA (that is, each
$N^i(\bfx)$ is summed from $0$ to $2^L-1$), where $\beta$, $\bfalpha$
and $\mu$ are Lagrange multipliers, and where
\[
M \equiv \sum_\bfx^{V} \sum_i^n N^i(\bfx),
\]
\[
{\bf P} \equiv \sum_\bfx^{V} \sum_i^n N^i(\bfx)\bfci
\]
and
\[
E \equiv \sum_\bfx^{V} \sum_i^n N^i(\bfx)\varepsilon_i
\]
are the total mass, momentum and energy, respectively, of all the
particles on a lattice of $V$ sites.  Thus, we have
\bgeas
{\cal Z}
 &=&
 \sum_{\{N\}}
  \exp
   \left[
    -\beta
    \sum_\bfx^{V}
    \sum_i^n
    \left(\varepsilon_i-\bfalpha\cdot\bfci-\mu\right)
    N^i(\bfx)
   \right]\\
 &=&
 \sum_{\{N\}}
  \prod_\bfx^{V}
  \prod_i^n
  \exp
   \left[
    -\beta
    \left(\varepsilon_i-\bfalpha\cdot\bfci-\mu\right)
    N^i(\bfx)
   \right]\\
 &=&
 \prod_\bfx^{V}
 \prod_i^n
 \sum_{k=0}^{2^L-1}
 \exp
  \left[
   -\beta
   \left(\varepsilon_i-\bfalpha\cdot\bfci-\mu\right)
   k
  \right]
 =
 \prod_\bfx^{V}
 \prod_i^n
 \sum_{k=0}^{2^L-1}
 (z^i)^k\\
 &=&
 \prod_\bfx^{V}
 \prod_i^n
 \left(
  \frac{1-(z^i)^{2^L}}{1-z^i}
 \right)
 =
 \left[
  \prod_i^n
  \left(
   \frac{1-(z^i)^{2^L}}{1-z^i}
  \right)
 \right]^{V},
\eeas
where we have defined the {\it fugacity}
\bge
z^i\equiv
 \exp
 \left[
  -\beta
  \left(\varepsilon_i-\bfalpha\cdot\bfci-\mu\right)
 \right].
\label{eq:fugdef}
\ee
The grand potential is then
\[
\Omega=-\frac{1}{\beta}\ln {\cal Z}
      =-\frac{V}{\beta}\sum_i^n
       \ln\left(\frac{1-(z^i)^{2^L}}{1-z^i}\right),
\]
so that
\[
\frac{\partial}{\partial\beta}\left(\beta\Omega\right) =
 -\frac{\partial}{\partial\beta}\ln {\cal Z} =
 V\sum_i^n
  \left(
   \frac{z^i}{1-z^i} -
   \frac{2^L(z^i)^{2^L}}{1-(z^i)^{2^L}}
  \right)
  \left(\varepsilon_i-\bfalpha\cdot\bfci-\mu\right).
\]
We identify the equilibrium distribution function
\bge
F_L(z)\equiv
   \frac{z}{1-z} -
   \frac{2^L(z)^{2^L}}{1-(z)^{2^L}},
\label{eq:eqdist}
\ee
which gives the mean number of particles moving in each lattice
direction.  Since this has a maximum of $2^L-1$, we also define the
{\it fractional occupation number}
\[
f_L(z)\equiv\frac{F_L(z)}{2^L-1}.
\]
Figure~\ref{fig:mop} shows $f_L(z)$ plotted against $z$ for several
values of $L$.

\begin{figure}
\epsfbox{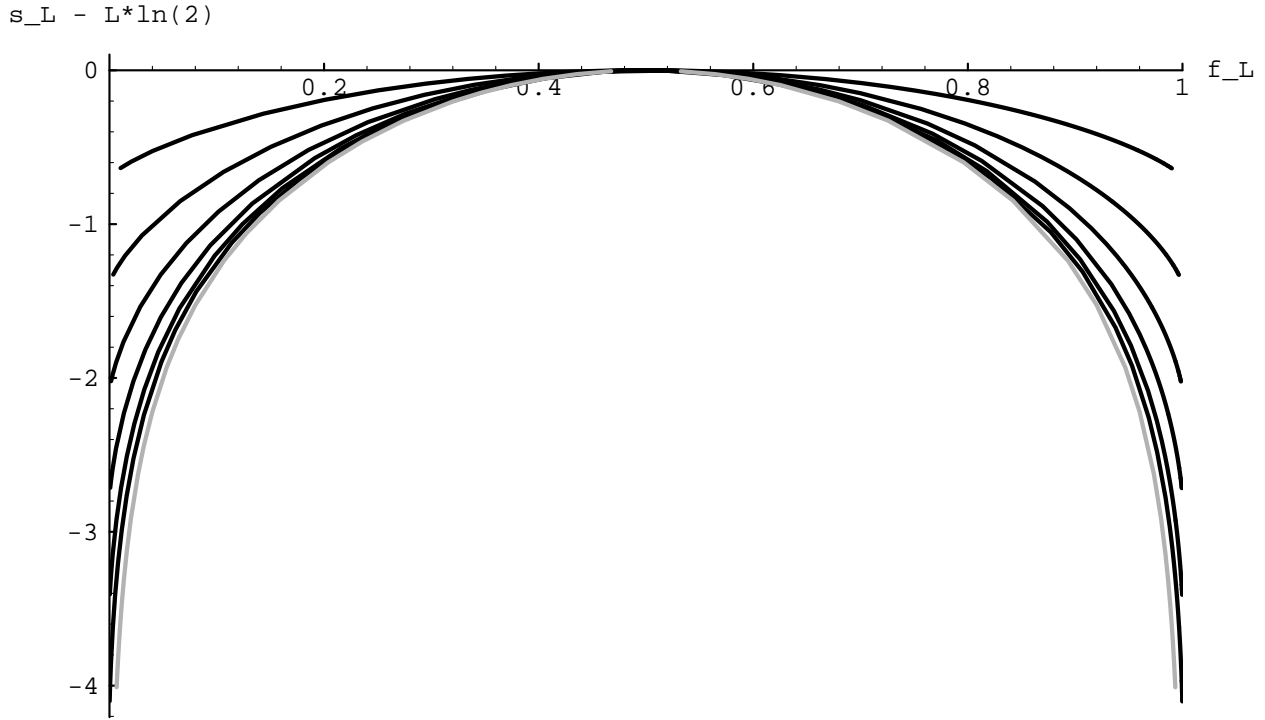}
\caption{Entropy excess $\Delta S_L$ versus fractional occupation number
  $f_L(z)$.  The black curves decrease with increasing $L$, and the gray
  curve is the limit as $L\rightarrow\infty$.}
\label{fig:ent}
\end{figure}

In terms of the equilibrium distribution function, we have
\[
\Omega+\beta\frac{\partial\Omega}{\partial\beta} =
\Omega-T\frac{\partial\Omega}{\partial T} =
 V\sum_i^n
  F_L(z^i)
  \left(\varepsilon_i-\bfalpha\cdot\bfci-\mu\right),
\]
where $T\equiv 1/\beta$ is the temperature.  It follows that
\[
\Omega=
 \langle H\rangle -
 \bfalpha\cdot\langle {\bf P}\rangle -
 \mu \langle M \rangle -
 T\langle S\rangle,
\]
where we have identified the average energy
\[
\langle H\rangle\equiv
 V\sum_i^n F_L(z^i)\varepsilon_i,
\]
the average momentum
\[
\langle {\bf P}\rangle\equiv
 V\sum_i^n F_L(z^i)\bfci,
\]
the average mass
\[
\langle M\rangle\equiv
 V\sum_i^n F_L(z^i),
\]
and the average entropy
\[
\langle S\rangle\equiv
 -\frac{\partial\Omega}{\partial T} =
 V\sum_i^n S_L(z^i).
\]
In the expression for the entropy we have defined the function
\bge
S_L(z)\equiv
   \ln\left(1-z^{2^L}\right) +
   \left(\frac{z^{2^L}}{1-z^{2^L}}\right)
    \ln\left(z^{2^L}\right) -
   \ln\left(1-z\right) -
   \left(\frac{z}{1-z}\right)
    \ln\left(z\right)
\label{eq:enta}
\ee
as the entropy per lattice direction.  Thus, in addition to the form for
the equilibrium distribution function, this analysis has provided us
with an expression for the entropy that is additive in the contributions
{}from each lattice direction.  In fact, it is straightforward to show
that $S_L \rightarrow L\ln 2$ in the limit of large $L$, corresponding
to a dominant contribution of $\ln 2$ per bit of state.  The excess
\bge
 \Delta S_L \equiv S_L - L\ln 2
 \label{eq:entb}
\ee
is then ${\cal O}(1)$ in $L$ and is plotted against the fractional
occupation number $f_L(z)$ in Fig.~\ref{fig:ent}.  This can be
interpreted as indicating that the bits are most random at half filling;
elsewhere, the entropy is lower than $L\ln 2$ per bit.

\section{Kinetic-Theoretical Treatment}

As an alternative to the preceding thermodynamic treatment of the
integer lattice gas, we can derive the principal results from a
kinetic-theoretical argument.  For example, to derive the form of the
equilibrium distribution function, Eq.~(\ref{eq:eqdist}), we can note
that the equilibrium distribution function {\it for each bit} must still
be Fermi-Dirac in form, since each individual bit is either occupied or
not.  Thus,
\[
N_0^{i,\ell}
 =
 \frac{1}{1+\exp\left[2^\ell\beta\left(\varepsilon_i-
                      \bfalpha\cdot\bfci-\mu\right)\right]},
\]
where the multipliers $\beta$, $\bfalpha$ and $\mu$ are determined in
terms of the mass, momentum and energy densities by their definitions,
Eqs.~(\ref{eq:masdef}), (\ref{eq:momdef}) and (\ref{eq:enedef}).  In
terms of the {\it fugacity}, Eq.~(\ref{eq:fugdef}), the above may be
written,
\bge
N_0^{i,\ell} = \frac{1}{1+(z^i)^{-2^\ell}}.
\label{eq:nodef}
\ee
The equilibrium distribution function for each direction is then given
by Eqs.~(\ref{eq:nidef}) and (\ref{eq:nodef}),
\[
N_0^i
 =
 \sum_{\ell=0}^{L-1} 2^\ell N_0^{i,\ell}
 =
 F_L(z^i),
\]
where we have defined the function
\bge
F_L(z)
 \equiv
 \sum_{\ell=0}^{L-1}\frac{2^\ell}{1+z^{-2^\ell}}.
\label{eq:fmdef}
\ee
In Appendix~\ref{sec:a1} we show that this sum is equal to the closed
form derived in the previous section,
\bge
F_L(z)
 =
 \frac{z}{1-z} - \frac{2^L z^{2^L}}{1-z^{2^L}}.
\label{eq:fmres}
\ee

\section{Form of the Hydrodynamic Equations}
\label{sec:form}

To derive the hydrodynamic equations, we first expand the equilibrium
distribution function in the Mach number.  Here and henceforth, we
specialize to the case of no internal energy, so that $\varepsilon_i=0$,
and we can absorb the multiplier $\beta$ into $\bfalpha$ and $\mu$.
Treating $\bfalpha$ as a small quantity, the fugacity can be written
\[
z^i = e^\mu\left(1+\bfalpha\cdot\bfci+
      \frac{1}{2}\bfalpha\bfalpha : \bfci\bfci
      \right) = z_0 + z_1^i + z_2^i,
\]
where the subscripts of
\bgeas
z_0   &\equiv& e^\mu \\
z_1^i &\equiv& z_0\bfalpha\cdot\bfci \\
z_2^i &\equiv& \frac{z_0}{2}\bfalpha\bfalpha : \bfci\bfci
\eeas
denote the order of the Mach number expansion, and we note that $z_0$ is
independent of the direction $i$.  It follows that
\[
N^i_0 = F_L(z^i) = F_L(z_0 + z_1^i + z_2^i).
\]
Taylor expanding, we get
\[
N^i_0 = F_L(z_0) + z_0 F_L^\prime (z_0)\bfalpha\cdot\bfci +
   \frac{1}{2}z_0\left[z_0 F_L^\prime (z_0)\right]^\prime
   \bfalpha\bfalpha : \bfci\bfci.
\]

To proceed, we must make some assumptions about the symmetries of the
lattice.  We demand that
\bge
\sum_{i=0}^{n-1}
 \bigotimes^k\bfci =
 A_k {\bf 1}_k
 \label{eq:latsym}
\ee
for $0\leq k\leq 4$, where $\otimes$ denotes the outer product, and
where ${\bf 1}_k$ is the completely symmetric and isotropic tensor of
rank $k$,
\[
{\bf 1}_0 = 1
\]
\[
({\bf 1}_1)_i = 0
\]
\[
({\bf 1}_2)_{ij} = \delta_{ij}
\]
\[
({\bf 1}_3)_{ijk} = 0
\]
\[
({\bf 1}_4)_{ijkl} = \delta_{ij}\delta_{kl} +
                     \delta_{ik}\delta_{jl} +
                     \delta_{il}\delta_{jk}.
\]
Note that Eq.~(\ref{eq:latsym}) {\it defines} the coefficients $A_k$ for
a given lattice.

We now demand that
\[
\rho = \sum_{i=0}^{n-1} N^i_0
     = A_0 F_L(z_0) +
       \frac{A_2\beta^2}{2} z_0\left[z_0 F_L^\prime (z_0)\right]^\prime,
\]
and
\[
\bfu = \sum_{i=0}^{n-1} \bfci N^i_0
     = A_2 z_0 F_L^\prime (z_0)\bfalpha.
\]
If we now let $z$ denote the solution to the equation
\bge
 \frac{\rho}{A_0}=F_L(z),
 \label{eq:dendef}
\ee
it follows that the difference between $z$ and $z_0$ is of second order
in the Mach number, so that we can solve for $\mu$ and $\bfalpha$.  We
find that
\[
\bfalpha = \frac{\bfu}{A_2 z F_L^\prime (z)},
\]
and that $\mu=\ln z_0$ where $z_0$ is the solution to the equation
\[
F_L(z_0)=\frac{\rho}{A_0} -
   \frac{z\left[z F_L^\prime (z)\right]^\prime}
        {2 A_0A_2\left[z F_L^\prime (z)\right]^2} u^2.
\]
Inserting these results into the distribution function, we find
\bge
N^i_0 =
 \frac{\rho}{A_0} +
 \frac{\bfu\cdot\bfci}{A_2} +
 \frac{z\left[z F_L^\prime (z)\right]^\prime}
      {2 A_2^2 \left[z F_L^\prime (z)\right]^2}
 \left(
  \bfci\bfci - \frac{A_2}{A_0}{\bf 1}_2
 \right) : \bfu\bfu,
\label{eq:nres}
\ee
where, again, $z$ is defined by $F_L (z)=\rho/A_0$.

The inviscid part of the stress tensor is then given by
\bgeas
\sum_{i=0}^n \bfci\bfci N^i_0
 &=&
 \frac{A_2}{A_0}\rho +
 \frac{z\left[z F_L^\prime (z)\right]^\prime}
      {2 A_2 \left[z F_L^\prime (z)\right]^2}
 \left(
      \frac{A_4}{A_2} {\bf 1}_4 - \frac{A_2}{A_0}{\bf 1}_2\otimes {\bf 1}_2
 \right) : \bfu\bfu \\
 &=&
 \left[
 \frac{A_2}{A_0}\rho +
 \frac{z\left[z F_L^\prime (z)\right]^\prime}
      {2 A_2 \left[z F_L^\prime (z)\right]^2}
 \left(
      \frac{A_4}{A_2} - \frac{A_2}{A_0}
 \right) u^2\right] {\bf 1}_2 +
 \frac{A_4 z\left[z F_L^\prime (z)\right]^\prime}
      {A_2^2 \left[z F_L^\prime (z)\right]^2} \bfu\bfu \\
 &=&
 P(\rho,u) {\bf 1}_2 + g(\rho)\frac{\bfu\bfu}{\rho},
\eeas
where we have identified the factor that multiplies the inertial term in
the Navier-Stokes equations,
\bge
g(\rho)
 =
 \frac{A_0 A_4 z F_L(z)\left[z F_L^\prime (z)\right]^\prime}
      {A_2^2 \left[z F_L^\prime (z)\right]^2},
 \label{eq:gdef}
\ee
and the equation of state,
\bge
P(\rho,u) =
 \frac{A_2}{A_0}\rho +
 \left(
   1 - \frac{A_2^2}{A_0 A_4}
 \right)
 g(\rho)
 \frac{u^2}{2\rho}.
 \label{eq:pdef}
\ee

Eqs.~(\ref{eq:dendef}), (\ref{eq:gdef}) and (\ref{eq:pdef}) are the
principal results of this section.  Eq.~(\ref{eq:dendef}) gives $\rho$
in terms of the parameter $z$.  Eq.~(\ref{eq:gdef}) then gives $g$ in
terms of $z$, so that Eqs.~(\ref{eq:dendef}) and (\ref{eq:gdef}) are a
pair of parametric algebraic equations for $g$ in terms of the density
$\rho$.  Finally, Eq.~(\ref{eq:pdef}) gives the equation of state for
$P$ in terms of $\rho$ and $\bfu$.  The coefficients $A_j$ that appear
in these equations are given in terms of the lattice vectors by the
conditions, Eq.~(\ref{eq:latsym}).

\section{Example: Bravais Lattice}

\begin{figure}
\epsfbox{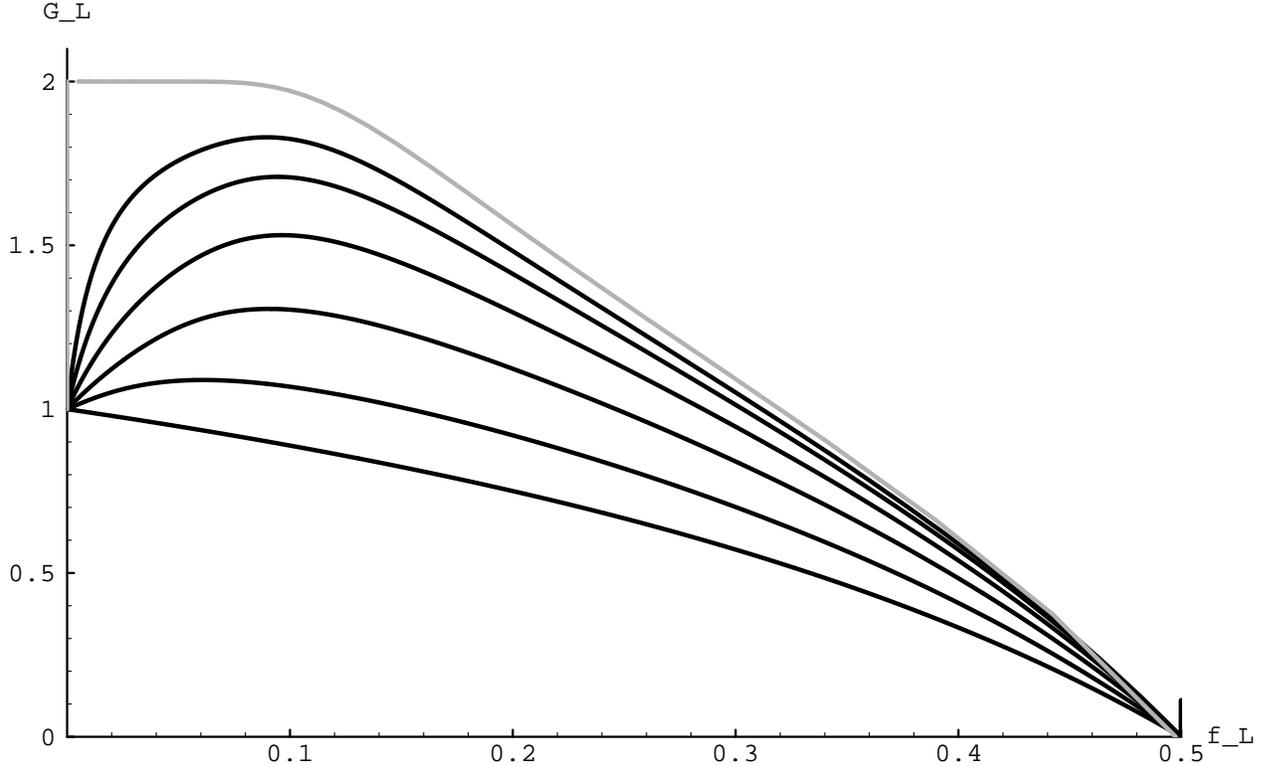}
\caption{$G_L$ versus $f$ for several values of $L$.  The black curves
represent $L$ values from 1 to 6, increasing upward, while the gray
curve is the limit as $L\rightarrow\infty$.}
\label{fig:gvsrho}
\end{figure}

As a concrete example of this formalism, we consider the case of a
regular Bravais lattice.  Examples of such lattices with the requisite
symmetry conditions, Eq.~(\ref{eq:latsym}), are the triangular lattice
in two dimensions~\cite{bib:fhp} and the face-centered hypercubic
lattice in four dimensions~\cite{bib:fchc}.  In addition to the $n$
directions corresponding to unit-speed particles, we include $n_r$ null
lattice vectors to accomodate rest particles.  In this situation,
\[
A_0
 = n + n_r
\]
\[
A_2
 = \frac{n}{D}
\]
and
\[
A_4
 = \frac{n}{D(D+2)},
\]
where $D$ is the number of dimensions.  Inserting these into
Eqs.~(\ref{eq:dendef}) through (\ref{eq:pdef}), we find
\[
\rho = \left( n + n_r \right) F_L(z),
\]
\[
g(\rho)
 =
 \left(
  \frac{D}{D+2}
 \right)
 \left(
  1+\frac{n_r}{n}
 \right)
 G_L(z),
\]
and
\[
P(\rho, u) =
 \frac{1}{D}
 \left(
  \frac{n}{n+n_r}
 \right)
 \left[\rho -
 \left(
  1 - \frac{D n_r}{2n}
 \right)
 g(\rho)
 \frac{u^2}{\rho}
 \right].
\]
Here we have defined the function,
\bge
G_L (z)\equiv
 \frac{z f_L(z) \left[ z f_L^\prime (z)\right]^\prime}
      { \left[ z f_L^\prime (z)\right]^2},
\label{eq:gdefbrav}
\ee
which we plot against the {\it fractional occupation number},
\[
f_L(z)\equiv
 \frac{\rho}{\left( 2^L-1\right) \left( n+n_r\right)}=
 \frac{F_L(z)}{2^L-1},
\]
for several different values of $L$ in Fig.~\ref{fig:gvsrho}.  For $L=1$
it is a straightforward exercise to show that we recover the well known
result~\cite{bib:fchc},
\[
G_1 (z) = \frac{1-2f}{1-f},
\]
which decreases monotonically from unity at $f=0$, to zero at
half-filling ($f=1/2$), after which it becomes negative.  For $L>1$, we
see that this decrease is no longer monotonic, since the slope at the
origin, $g'(0)$, is positive.  Thus, for $L>1$, the function $g$ has a
maximum for some $0<f<1/2$.  The location of this maximum approaches
$f=0$ as $L\rightarrow\infty$.  (The limit of infinite integers, i.e.,
$L\rightarrow\infty$ is discussed in Appendix~\ref{sec:a3}, and is shown
as a shaded curve in Fig.~\ref{fig:gvsrho}.)

Galilean invariance is achieved when $g=1$, or
\[
G_L = \left( 1+\frac{2}{D} \right)
      \left( \frac{n}{n+n_r} \right).
\]
If the quantity $(1+2/D)n/(n+n_r)$ is greater than the maximum value of
$G_L$, then galilean invariance is impossible for those values of $D$,
$n$, $n_r$, and $L$; if it is less than this maximum, then there are two
densities at which galilean invariance is achieved.  Some of these
values are tabulated for the FHP and FCHC lattice gases in
Fig.~\ref{fig:froot}.
\begin{figure}
\begin{center}
\begin{tabular}{|c|c|ll|}
\hline
\multicolumn{4}{|c|}{FHP Lattice Gas ($D=2$, $n=6$)} \\
\hline
\hline
$n_r$ & $L$ & Low-Density Root & High-Density Root \\
\hline
0 & $\infty$ & 0.0        & 0.0      \\
\hline
1 & $6$      & 0.0396831  & 0.143848 \\
  & $\infty$ & 0.0        & 0.168451 \\
\hline
2 & $4$      & 0.0704358  & 0.126560 \\
  & $5$      & 0.0322583  & 0.177356 \\
  & $6$      & 0.0158730  & 0.195949 \\
  & $\infty$ & 0.0        & 0.212636 \\
\hline
4 & $3$      & 0.0362392  & 0.167555 \\
  & $4$      & 0.0166667  & 0.228582 \\
  & $5$      & 0.0080645  & 0.253770 \\
  & $6$      & 0.0039683  & 0.265426 \\
  & $\infty$ & 0.0        & 0.276535 \\
\hline
\end{tabular}
\end{center}
\begin{center}
\begin{tabular}{|c|c|ll|}
\hline
\multicolumn{4}{|c|}{FCHC Lattice Gas ($D=4$, $n=24$)}\\
\hline
\hline
$n_r$ & $L$ & Low-Density Root & High-Density Root \\
\hline
0 & $4$      & 0.0704358  & 0.126560 \\
  & $5$      & 0.0322583  & 0.177356 \\
  & $6$      & 0.0158730  & 0.195949 \\
  & $\infty$ & 0.0        & 0.212636 \\
\hline
1 & $4$      & 0.0528917  & 0.152419 \\
  & $5$      & 0.0253456  & 0.193030 \\
  & $6$      & 0.0124717  & 0.209795 \\
  & $\infty$ & 0.0        & 0.225163 \\
\hline
2 & $4$      & 0.0417410  & 0.171769 \\
  & $5$      & 0.0201613  & 0.207212 \\
  & $6$      & 0.0099206  & 0.222576 \\
  & $\infty$ & 0.0        & 0.236849 \\
\hline
4 & $3$      & 0.0654937  & 0.120009 \\
  & $4$      & 0.0266677  & 0.203001 \\
  & $5$      & 0.0129032  & 0.232239 \\
  & $6$      & 0.0063492  & 0.245488 \\
  & $\infty$ & 0.0        & 0.257994 \\
\hline
\end{tabular}
\end{center}
\caption{Values of $f\in (0,1/2)$ such that $g=1$}
\label{fig:froot}
\end{figure}

\section{Viscosity}

To compute the viscosity of a ILGA in the Boltzmann molecular chaos
approximation~\cite{bib:fchc}, we consider its ensemble-averaged
collision operator, $\Omega^{i,\ell}$.  This quantity is the ensemble
average of the increase in bit $\ell$ in direction $i$ due to
collisions.  It is given by
\[
\Omega^{i,\ell} =
 \sum_{s,s'} A(s\rightarrow s') (s^{\prime i,\ell} - s^{i,\ell})
 {\cal P}(s),
\]
where ${\cal P}(s)$ is the probability that the incoming state is $s$,
$A(s\rightarrow s')$ is the probability that the collision process takes
incoming state $s$ to outgoing state $s'$, and $s^{i,\ell}$ is the value
of bit $\ell$ in direction $i$ in incoming state $s$ (and likewise for
outgoing state $s'$).  In the Boltzmann approximation, the probability
of a state $s$ is the product of the corresponding fractional occupation
numbers, or their complements,
\[
{\cal P}(s) =
 \prod_{k'=0}^{n-1}\prod_{\jmath'=0}^{L-1}
 \left( N^{k',\jmath'} \right)^{s^{k',\jmath'}}
 \left( 1 - N^{k',\jmath'} \right)^{1-s^{k',\jmath'}}.
\]
To get the total increase of particles in direction $i$, we take the sum
\[
\Omega^i \equiv\sum_{\ell=0}^{L-1} 2^\ell \Omega^{i,\ell} =
 \sum_{s,s'} A(s\rightarrow s') (s^{\prime i} - s^{i})
 \prod_{k'=0}^{n-1}\prod_{\jmath'=0}^{L-1}
 \left( N^{k',\jmath'} \right)^{s^{k',\jmath'}}
 \left( 1 - N^{k',\jmath'} \right)^{1-s^{k',\jmath'}},
\]
where
\[
 s^i\equiv\sum_{\ell=0}^{L-1} 2^\ell s^{i,\ell}
\]
is the total number of particles in direction $i$ in state $s$ (and
likewise for $s'$).

To compute the viscosity, we must form the Jacobian matrix of the
collision operator.  Direct calculation yields
\[
\frac{\partial\Omega^{i,\ell}}{\partial N^{k,\jmath}} =
 \sum_{s,s'} A(s\rightarrow s') (s^{\prime i} - s^{i})
 {\cal P}(s)
 \frac{\left(s^{k,\jmath} - N^{k,\jmath}\right)}
 {N^{k,\jmath} \left( 1-N^{k,\jmath}\right)}.
\]
We would like to evaluate this Jacobian at the equilibrium given by
Eq.~(\ref{eq:nodef}),
\[
 N_0^{k,\jmath} = \frac{1}{1+(z^k)^{-2^\jmath}}.
\]
Taking the derivative of this equation with respect to the fugacity,
\[
 \frac{\partial N_0^{k,\jmath}}{\partial z^k} =
 \frac{2^\jmath (z^k)^{-2^\jmath-1}}
 {\left[1+(z^k)^{-2^\jmath}\right]^2} =
 \frac{2^\jmath N_0^{k,\jmath} \left( 1-N_0^{k,\jmath}\right)}
 {z^k},
\]
we can use the chain rule to get the integer version of the
Jacobian of the collision operator at equilibrium,
\bgea
\mixten{J}{i}{k} &\equiv&
 \left. \frac{\partial\Omega^i}{\partial N^k} \right|_0
 \nonumber\\
 &=&
 \sum_{\ell,\jmath=0}^{L-1} 2^\ell \;
 \left.
 \frac{\partial\Omega^{i,\ell}}
      {\partial N^{k,\jmath}}
 \right|_0
 \left.
 \frac{\partial N^{k,\jmath}/\partial z^k}
      {\partial N^k/\partial z^k}
 \right|_0
 \nonumber\\
 &=&
 \sum_{s,s'} A(s\rightarrow s') (s^{\prime i} - s^{i})
 {\cal P}_0 (s)
 \frac{\left(s^{k} - N^{k}\right)}
 {z^k F_L^\prime (z^k)}.
\label{eq:copt}
\eea
In fact, we need this result only in the limit of zero Mach number, so
we can use the lowest order expression for the fugacity, $z^k=z$ (see
Sec.~\ref{sec:form}), which is independent of the index $k$.  We find
that the zero Mach number limit of the Boltzmann probability of state
$s$ is given by
\[
{\cal P}_0(s) =
 \prod_{k'=0}^{n-1}\prod_{\jmath'=0}^{L-1}
 \left( N_0^{k',\jmath'} \right)^{s^{k',\jmath'}}
 \left( 1 - N_0^{k',\jmath'} \right)^{1-s^{k',\jmath'}}
 =
 \prod_{\jmath'=0}^{L-1}
 \left(
 \frac{1}{1+z^{-2^{\jmath'}}}
 \right)^{p_{\jmath'} (s)}
 \left(
 \frac{z^{-2^{\jmath'}}}{1+z^{-2^{\jmath'}}}
 \right)^{n-p_{\jmath'} (s)},
\]
where
\[
p_\jmath (s)\equiv\sum_{k=0}^{n-1} s^{k,\jmath}
\]
is the total number of populated bits in the $\jmath$th binary digit.
It follows that
\[
{\cal P}(s) =
 \left[
 \prod_{\jmath=0}^{L-1}
  \left(
  1 + z^{2^\jmath}
  \right)^{-1}
 \right]^n
 \left(
 \prod_{\jmath=0}^{L-1}
 z^{2^\jmath p_\jmath (s)}
 \right) =
 \left(
 \frac{1-z}{1-z^{2^L}}
 \right)^n
 z^{p(s)},
\]
where
\[
p(s)\equiv\sum_{\jmath=0}^{L-1} 2^\jmath p_\jmath (s)
\]
is the total number of particles present in state $s$.

Inserting this result into the expression, Eq.~(\ref{eq:copt}), for the
collision operator, we obtain
\[
\mixten{J}{i}{k} =
 \frac{1}{z F_L^\prime (z)}
 \left(
  \frac{1-z}{1-z^{2^L}}
 \right)^n
 \sum_{s,s'}
 A(s\rightarrow s')
 \left( s^{\prime i} - s^i\right)
 \left[ s^k - F_L (z) \right]
 z^{p(s)}.
\]
As a consequence of conservation of probability,
Eq.~(\ref{eq:consprob}), and semidetailed balance, Eq.~(\ref{eq:sdb}),
it follows that the second term in square brackets vanishes, so we
finally get
\[
\mixten{J}{i}{k} =
 \frac{1}{z F_L^\prime (z)}
 \left(
  \frac{1-z}{1-z^{2^L}}
 \right)^n
 \sum_{s,s'}
 A(s\rightarrow s')
 \left( s^{\prime i} - s^i\right) s^k
 z^{p(s)}.
\]

At first order in Knudsen number, the kinetic equation
is~\cite{bib:long}
\[
\bfci\cdot\bfnabla N_0^i = \mixten{J}{i}{j} N_1^j,
\]
where there is an understood summation over $j$.  The only part of the
left-hand side that contributes to the viscosity comes from the second
term on the right-hand side of Eq.~(\ref{eq:nres}), whence
\[
\mixten{J}{i}{j} N_1^j = \frac{1}{A_2}\bfci\bfci : \bfnabla\bfu.
\]
Now, $J$ is a singular matrix; it has a null eigenvector corresponding
to each hydrodynamic mode of the system.  These null eigenvectors span
what we shall call the {\it hydrodynamic subspace} of the system.  The
complement of this subspace is called the {\it kinetic subspace}, and is
spanned by the kinetic modes with nonzero (negative) eigenvalue.  If we
restrict our attention to the kinetic subspace, then we can form the
pseudoinverse of $J$, denoted by $J^{-1}$, in terms of which we may
write
\[
N_1^i = \frac{1}{A_2}\mixten{\left(J^{-1}\right)}{i}{j}
 \bfcj\bfcj : \bfnabla\bfu.
\]
The conservation law for momentum then contains the term
\bgeas
\lefteqn{
\sum_i
 \left(
  \bfci\bfci\cdot\bfnabla N_1^i +
  \smallhalf\bfci\bfci\bfci : \bfnabla\bfnabla N_0^i +
  \cdots
 \right) } \\
 &=&
 \bfnabla\cdot
  \left\{
   \frac{1}{A_2}
   \left[
    \sum_{i,j}
    \bfci\bfci
    \mixten{\left(J^{-1}\right)}{i}{j}
    \bfcj\bfcj +
    \smallhalf
    \sum_i
    \bfci\bfci\bfci\bfci
   \right] :
  \left(\bfnabla\bfu\right)
 \right\} + \cdots.
\eeas
We note that $J^{-1}$ is diagonalized and degenerate in the subspace
spanned by the $n$ outer products of the lattice vectors with
themselves; that is
\bge
 \sum_j\mixten{\left(J^{-1}\right)}{i}{j}\bfcj\bfcj = -\lambda\bfci\bfci,
 \label{eq:ansatz}
\ee
where $\lambda$ is a scalar, whence the above term in the momentum
conservation equation becomes
\[
 \bfnabla\cdot
 \left[
  \frac{A_4}{A_2}
  \left(
   -\lambda+\smallhalf
  \right)
  {\bf 1}_4 :
  \left(\bfnabla\bfu\right)
 \right] + \cdots
 =
 \bfnabla\cdot
 \left[
  \frac{A_4}{A_2}
  \left(
  -\lambda+\smallhalf
  \right)
  \bfnabla\bfu
 \right],
\]
{}from which we identify the kinematic viscosity,
\[
\nu = \frac{A_4}{A_2}\left(\lambda-\smallhalf\right).
\]
The quantity $\lambda$ is then determined by taking the double spatial
dot product of $\bfci\bfci$ on both sides of Eq.~(\ref{eq:ansatz}), and
summing over $i$ to get
\[
n = -\lambda\sum_{i,j}\mixten{J}{i}{j}\left(\bfci\cdot\bfcj\right)^2,
\]
whence
\[
\frac{1}{\lambda} =
 \frac{-1}{nzF_L^\prime (z)}
 \left(
  \frac{1-z}{1-z^{2^L}}
 \right)^n
 \sum_{i,j}
 \sum_{s,s'}
 A(s\rightarrow s')
 \left(s^{\prime i} -s^i\right)
 z^{p(s)} s^j
 \left(\bfci\cdot\bfcj\right)^2,
\]
where $f=f_L(z)$ determines the parameter $z$ in terms of the fractional
occupation number.  This result is easily seen to reduce to that of
H\'{e}non~\cite{bib:henonvis} when $L=1$.

\begin{figure}
\epsfxsize=400pt\epsfbox{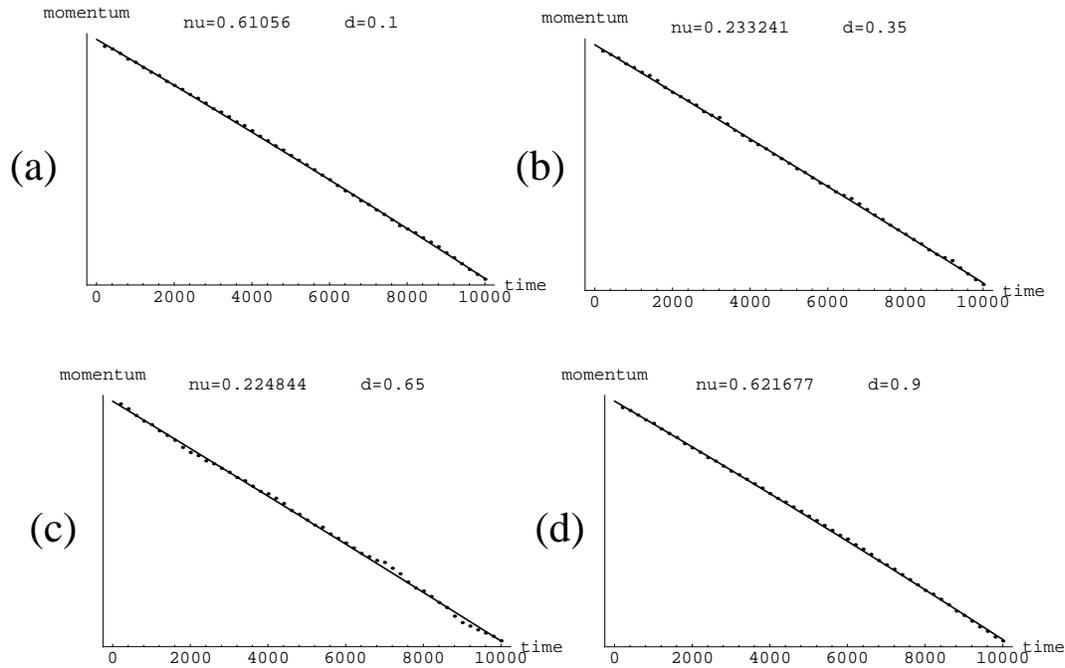}
\caption{Time decay of shear-wave amplitude}
\label{fig:decay}
\end{figure}

\begin{figure}
\epsfxsize=400pt\epsfbox{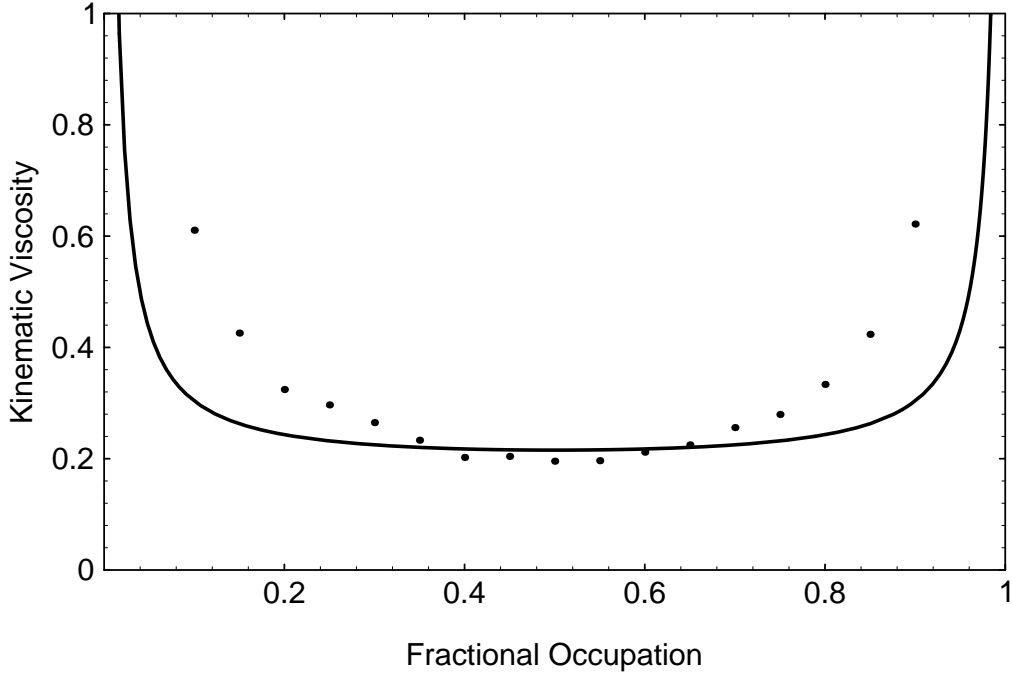}
\caption{Viscosity versus $f$ for $L=2$.}
\label{fig:vvsrho}
\end{figure}

We computed the viscosity of an $L=2$ lattice gas in two dimensions
($D=2$) by measuring the decay of a shear wave in periodic geometry.  We
used a lattice of size $512\times 512$ on a CAM-8 Cellular-Automata
Machine~\cite{bib:cam}.  The probabilistic collision procedure used
obeyed semi-detailed balance, with each outgoing state allowed by the
conservation laws sampled with equal probability.  Fig.~\ref{fig:decay}
shows the decay of the shear wave amplitude to be exponential in nature,
as is appropriate for Navier-Stokes evolution.  The time constant of the
exponential then determines the viscosity, which is plotted as a
function of density in Fig.~\ref{fig:vvsrho}, along with the curve
predicted by the theory given above.
While the agreement with theory is good at intermediate values of the
fractional occupation number near half filling, we note that it is
seriously in error at low (and high) fractional occupation numbers.  At
present, we attribute this discrepency to deviations from the Boltzmann
molecular chaos approximation, and we plan to investigate them using
kinetic ring theory~\cite{bib:long} in a forthcoming publication.

\section{Statistical Noise}

Finally, we consider the statistical noise of the ILGA model.  With the
maximum number of particles per direction increasing as $2^L$, one might
naively expect the noise level to decrease with $L$ as $1/\sqrt{2^L}\sim
2^{-L/2}$.  Unfortunately, as we shall show, this expectation is not
realized, due to the extremely narrow dynamic range of the fugacity for
large $L$.  This is best seen in Fig.~\ref{fig:mop}, in which the
effective width of the function $f_L(z)$ near $z=\smallhalf$ decreases
like $2^{-L}$, making for a subtle limiting process that is discussed in
Appendix~\ref{sec:a3}.

Let $n^{i,\ell}(\bfx,t)$ be the precise value of bit $\ell$ in direction
$i$ at lattice site $\bfx$ at time $t$.  The ensemble-average of this
quantity is $N^{i,\ell}$, as used in the text of the paper.  The mean
number of particles in a (space-time) block of $N$ sites is then
\[
{\cal F}_1 = 
 \sum_{(\bfx,t)}^N
 \sum_i^n
 \sum_{\ell=0}^{L-1}
 2^\ell
 \left\langle n^{i,\ell}(\bfx,t)\right\rangle =
 nNF_L(z),
\]
where the angle brackets denote the ensemble average.

\begin{figure}
\epsfbox{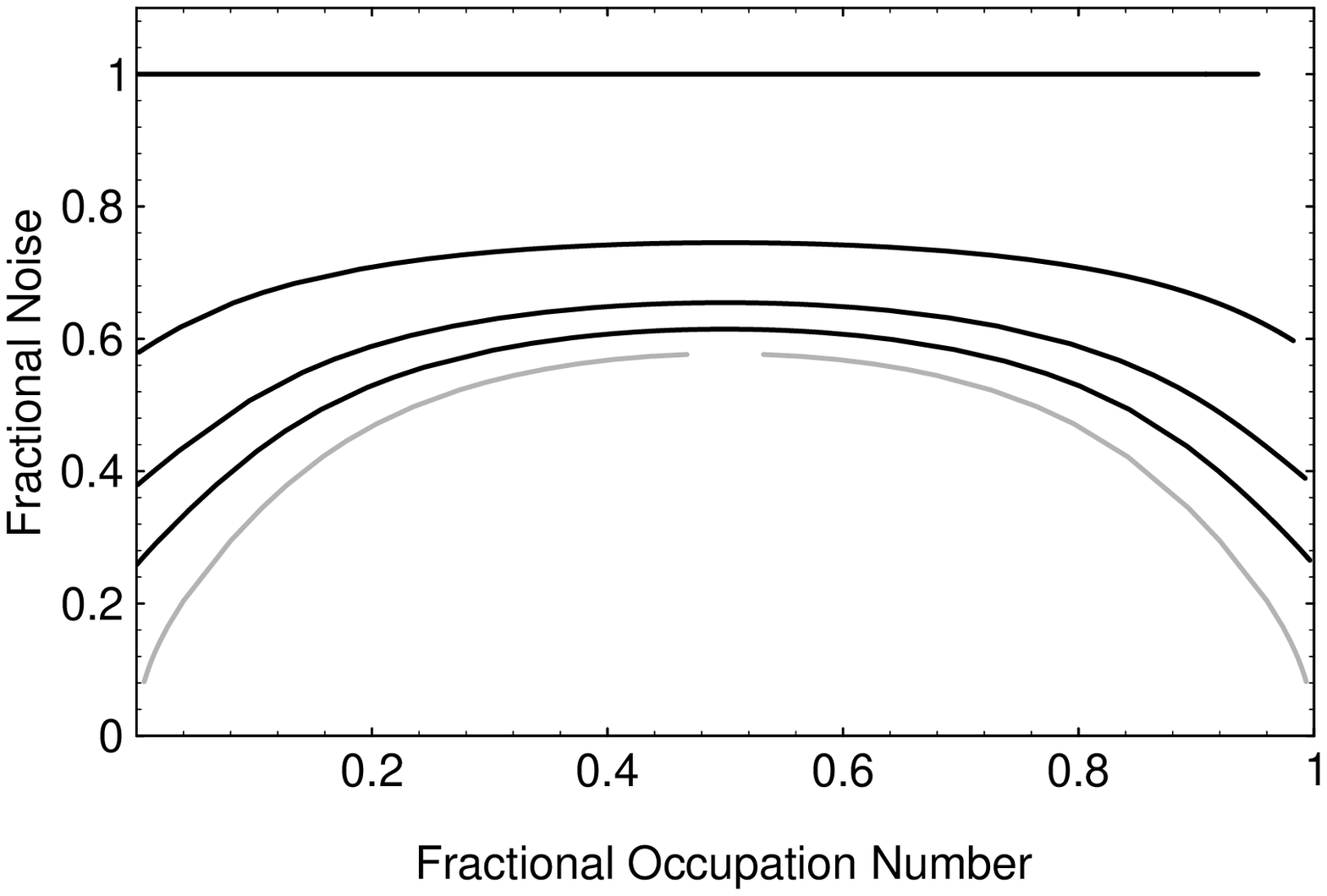}
\caption{$\Delta {\cal F}_L$ versus $f$ for several values of $L$.  The
  black curves represent $L$ values from 1 to 6, increasing downward,
  while the gray curve is the limit as $L\rightarrow\infty$.}
\label{fig:nois}
\end{figure}

The mean square of the number of particles in this block of sites is
then
\[
{\cal F}_2 =
 \sum_{(\bfx,t)}^N
 \sum_{(\bfx',t')}^N
 \sum_i^n
 \sum_{i'}^n
 \sum_{\ell=0}^{L-1}
 \sum_{\ell'=0}^{L-1}
 2^{\ell+\ell'}
 \left\langle n^{i,\ell}(\bfx,t)n^{i',\ell'}(\bfx',t')\right\rangle.
\]
The bits are either zero or one, and in the Boltzmann molecular chaos
approximation different bits are uncorrelated.  It follows that
\[
\left\langle n^{i,\ell}(\bfx,t)n^{i',\ell'}(\bfx',t')\right\rangle =
 \left\langle n^{i,\ell}(\bfx,t)\right\rangle
 \left\langle n^{i',\ell'}(\bfx',t')\right\rangle +
 \delta_{\bfx,\bfx'}
 \delta_{t,t'}
 \delta_{i,i'}
 \delta_{\ell,\ell'}
 \left\langle n^{i,\ell}(\bfx,t)\right\rangle
 \left(
  1-\left\langle n^{i,\ell}(\bfx,t)\right\rangle
 \right),
\]
whence
\[
{\cal F}_2 =
 {\cal F}_1^2 + nN\sum_\ell^{L-1}
 \frac{2^{2\ell} z^{2^\ell}}
      {\left(1+z^{2^\ell}\right)^2} =
 {\cal F}_1^2 + nNzF_L^\prime (z).
\]
It follows that the standard deviation of the number of particles in the
block is $\sqrt{{\cal F}_2-{\cal F}_1^2}$.  To define a fractional
noise, we could divide this by the mean number of particles, ${\cal
F}_1$, but it preserves particle-hole symmetry if we instead divide it
by the square root of the product of the mean number of particles and
the mean number of holes, thus
\bge
\Delta {\cal F}\equiv
 \sqrt{\frac{{\cal F}_2-{\cal F}_1^2}
      {{\cal F}_1\left[nN\left(2^L-1\right)-{\cal F}_1\right]}} =
 \frac{1}{\sqrt{nN\left(2^L-1\right)}}
 \sqrt{\frac{zf_L^\prime (z)}
      {f_L(z)\left[1-f_L(z)\right]}}.
\label{eq:nois}
\ee
This appears to decrease exponentially with $L$, but it must be noted
that the logarithmic derivative of $f_L(z)$ goes as $2^L$ at
$z=\smallhalf$.  Since, for fixed fractional occupation number $f_L$,
$z$ tends to $\smallhalf$ as $L$ tends to infinity, we see that $\Delta
{\cal F}$ is order unity in $L$.  Thus, the fractional noise does
decrease with $L$, but not as rapidly as one might hope.  It is plotted
for several different values of $L$ in Fig.~\ref{fig:nois}.

\section{Sampling Procedure}

Finally, we consider some practical considerations concerning the
computer implementation of the ILGA model.  Since each site has $nL$
bits, and therefore $2^{nL}$ possible states, and since the most popular
lattices with the requisite isotropy properties have $n=6$ and $n=24$,
it is clear that the brute-force approach in which a lookup table is
used to store the collision outcome states will not be feasible for $L$
much greater than unity.

For this reason, we propose another sampling scheme for the outgoing
states.  Though the method we propose is completely general, we
illustrate it for the two dimensional integer lattice gas on a
triangular grid ($n=6$).  Let ${\bf n}$ be an integer-valued column
$n$-vector whose components are the particle occupation numbers in each
of the six directions.

Let us suppose that the mass and the two components of momentum are the
only conserved quantities.  Since these conserved quantitites are linear
in the particle occupation numbers, each of them correspond to a row
vector, whose inner product with ${\bf n}$ yields the conserved quantity
in question.  Thus, corresponding to the mass we have the row vector
\[
{\bf q}_1 =
 \left(
  \begin{array}{rrrrrr}
   1 & 1 & 1 & 1 & 1 & 1
  \end{array}
 \right),
\]
corresponding to the $x$-momentum (multiplied by a factor of $2$), we
have
\[
{\bf q}_2 =
 \left(
  \begin{array}{rrrrrr}
   2 & 1 &-1 &-2 &-1 & 1
  \end{array}
 \right),
\]
and corresponding to the $y$-momentum (multiplied by a factor of
$2/\sqrt{3}$), we have
\[
{\bf q}_3 =
 \left(
  \begin{array}{rrrrrr}
   0 & 1 & 1 & 0 &-1 &-1
  \end{array}
 \right).
\]
In fact, these row vectors are precisely the {\it hydrodynamic
eigenvectors}, mentioned in our derivation of the viscosity; that is,
\[
\mixten{J}{i}{k} ({\bf q}_1)^k =
\mixten{J}{i}{k} ({\bf q}_2)^k =
\mixten{J}{i}{k} ({\bf q}_3)^k = 0.
\]
It is clear that these can always be chosen to be mutually orthogonal,
without loss of generality.  Using the Gram-Schmidt procedure, it is
then possible to find three vectors spanning the {\it kinetic subspace},
orthogonal to the above; e.g.,
\[
{\bf q}_4 =
 \left(
  \begin{array}{rrrrrr}
   2 &-1 &-1 & 2 &-1 &-1
  \end{array}
 \right),
\]
\[
{\bf q}_5 =
 \left(
  \begin{array}{rrrrrr}
   1 &-1 & 1 &-1 & 1 &-1
  \end{array}
 \right),
\]
and
\[
{\bf q}_6 =
 \left(
  \begin{array}{rrrrrr}
   0 & 1 &-1 & 0 & 1 &-1
  \end{array}
 \right).
\]

Now the collision process takes state ${\bf n}$ to state ${\bf n}'$.
Since it cannot change the values of the conserved quantities, it
follows that the difference between ${\bf n}'$ and ${\bf n}$ must be a
linear combination of kinetic eigenvectors.  That is,
\[
{\bf n}' = {\bf n} +
 \alpha_4 {\bf q}_4^T + \alpha_5 {\bf q}_5^T + \alpha_6 {\bf q}_6^T,
\]
where the $\alpha$'s are integer constants, and where the superscript
$T$ denotes ``transpose.''  Thus, writing out components, we have
\[
{\bf n}' =
 \left(
  \begin{array}{l}
  n_1 + 2\alpha_4 + \alpha_5 \\
  n_2 - \alpha_4 - \alpha_5 + \alpha_6 \\
  n_3 - \alpha_4 + \alpha_5 - \alpha_6 \\
  n_4 + 2\alpha_4 - \alpha_5 \\
  n_5 - \alpha_4 + \alpha_5 + \alpha_6 \\
  n_6 - \alpha_4 - \alpha_5 - \alpha_6
  \end{array}
 \right).
\]
Since the components of ${\bf n}'$ must all be between $0$ and $2^L-1$,
inclusive, we derive the following six inequality constraints:
\bgeas
  0 & \leq & n_1 + 2\alpha_4 + \alpha_5 \leq 2^L-1 \\
  0 & \leq & n_2 - \alpha_4 - \alpha_5 + \alpha_6 \leq 2^L-1 \\
  0 & \leq & n_3 - \alpha_4 + \alpha_5 - \alpha_6 \leq 2^L-1 \\
  0 & \leq & n_4 + 2\alpha_4 - \alpha_5 \leq 2^L-1 \\
  0 & \leq & n_5 - \alpha_4 + \alpha_5 + \alpha_6 \leq 2^L-1 \\
  0 & \leq & n_6 - \alpha_4 - \alpha_5 - \alpha_6\leq 2^L-1.
\eeas
These inequality constraints define a polytope in the three dimensional
space of allowed values of $\alpha_4$, $\alpha_5$ and $\alpha_6$.  We
know that this polytope must exist and contain the origin in that space,
since $\alpha_4 = \alpha_5 = \alpha_6 = 0$, corresponding to the
``trivial collision'' in which the occupation numbers do not change
their values, will always satisfy the constraints.

The collision process is then specified by a strategy for sampling
points from this polytope.  One viable strategy which certainly
satisfies semidetailed balance is to sample the points within this
polytope uniformly.  It is possible, though tedious, to derive a
closed-form algorithm to do this, based on the above constraints.
Alternatively, with some loss of efficiency, one can simply bound the
polytope and use a rejection sampling scheme.  Details of this procedure
will be provided in a forthcoming publication~\cite{bib:forthcom}.

\section{Conclusions}

We have generalized the hydrodynamic lattice gas model to include
integer numbers of particles moving in each direction at each site.  We
have presented the thermodynamics and kinetic theory of this generalized
Integer Lattice Gas (ILGA) model, including closed-form (or parametric
algebraic) equations for the equilibrium distribution function, the
entropy, the equation of state, the non-galilean factor in the inertial
term of the fluid equations, and the statistical noise.  We have thereby
shown that the ILGA model allows for the attainment of galilean
invariance, and a reduction in the kinematic viscosity and the
statistical noise.  In future publications, we shall show that this
generalization also allows for more straightforward inclusion of
interparticle interactions than the usual binary model.

\section*{Acknowledgements}

We are grateful to Xiaowen Shan and Harris Gilliam for useful
discussions and computer simulations that contributed to this study.
This work was supported in part by the Mathematical and Computational
Sciences Directorate of the Air Force Office of Scientific Research,
Initiative 2304CP.  Two of us (BMB and FJA) were supported in part by
IPA agreements with Phillips Laboratory.  BMB was also supported by
AFOSR grant number F49620-95-1-0285.

\appendix

\clearpage
\section{Closed-Form Expression for $F_L(z)$}
\label{sec:a1}
In this Appendix, we prove Eq.~(\ref{eq:fmres}), where $F_L(z)$ is
defined by Eq.~(\ref{eq:fmdef}).  Using mathematical induction, we first
note that the statement of the theorem is true for $L=1$:
\bgeas
F_1(z)
 &\equiv&
 \sum_{\ell=0}^{0}\frac{2^\ell}{1+z^{-2^\ell}}
 =
 \frac{1}{1+z^{-1}}
 =
 \frac{z}{1+z}
 =
 \frac{z(1-z)}{1-z^2}\\
 &=&
 \frac{z(1+z)}{1-z^2}-\frac{2z^2}{1-z^2}
 =
 \frac{z}{1-z} - \frac{2z^2}{1-z^2}.
\eeas
Next, we assume the truth of the statement for $L=K$:
\[
F_K(z)
 \equiv
 \sum_{\ell=0}^{K-1}\frac{2^\ell}{1+z^{-2^\ell}}
 =
 \frac{z}{1-z} - \frac{2^K z^{2^K}}{1-z^{2^K}}.
\]
It follows that
\bgeas
F_{K+1}(z)
 &\equiv&
 \sum_{\ell=0}^{K}\frac{2^\ell}{1+z^{-2^\ell}}
 =
 F_K(z) + \frac{2^K}{1+z^{-2^K}}\\
 &=&
 \frac{z}{1-z} -
 \frac{2^K z^{2^K}}{1-z^{2^K}} +
 \frac{2^K}{1+z^{-2^K}}
 =
 \frac{z}{1-z} +
 2^K
 \left(
 \frac{1}{1-z^{-2^K}} +
 \frac{1}{1+z^{-2^K}}
 \right)\\
 &=&
 \frac{z}{1-z} +
 2^K
 \frac{2}{1-\left(z^{-2^K}\right)^2}
 =
 \frac{z}{1-z} +
 \frac{2^{K+1}}{1-z^{-2^{K+1}}}\\
 &=&
 \frac{z}{1-z} -
 \frac{2^{K+1} z^{2^{K+1}}}{1-z^{2^{K+1}}},
\eeas
and we have thereby proven the theorem for all $K$.

Alternatively, we may simply note that the summation can be written in
the telescoping form
\[
F_L(z)
 \equiv
 \sum_{\ell=0}^{L-1}\frac{2^\ell}{1+z^{-2^\ell}}
 =
 \sum_{\ell=0}^{L-1}
  \left(
   \frac{2^\ell z^{2^\ell}}{1-z^{2^\ell}} -
   \frac{2^{\ell+1} z^{2^{\ell+1}}}{1-z^{2^{\ell+1}}}
  \right),
\]
{}from which the result follows immediately.

\clearpage
\section{The Infinite Integer Limit}
\label{sec:a3}

To consider the limit of infinite integers,
$L\rightarrow\infty$, we first note that the fractional occupation
number,
\bge
f_L(z)
 =
 \frac{F_L(z)}{2^L-1}
 =
 \frac{1}{2^L-1}
 \left(
  \frac{z}{1-z} -
  \frac{2^L z^{2^L}}{1-z^{2^L}}
 \right),
\label{eq:focdef}
\ee
has the limiting behavior
\[
\lim_{z\rightarrow 0} f_L(z) = 0
\]
\[
\lim_{z\rightarrow 1} f_L(z) = \mbox{{\scriptsize $\frac{1}{2}$}}
\]
\[
\lim_{z\rightarrow\infty} f_L(z) = 1
\]
for all $L$; here we have used L'H\^{o}pital's rule to establish the
result for $z\rightarrow 1$.  Referring to Figure~\ref{fig:mop}, we note
that the function $f_L(z)$ becomes increasingly like a step at $z=1$ as
$L\rightarrow\infty$.  To verify this, we note that the width of the
gradient there can be estimated by
\[
\lim_{z\rightarrow 1}
 \frac{f_L(z)}{f_L^\prime (z)} =
 \frac{6}{2^L + 1},
\]
which clearly goes to zero as $L\rightarrow\infty$; once again we have
used L'H\^{o}pital's rule to establish this result.

The approach to a step function means that the entire range of
fractional occupation numbers is parametrized by values of $z$ within
order $2^{-L}$ from $1$, as $L\rightarrow\infty$.  That being the case,
we write
\bge
z = 1+\frac{y}{2^L},
\label{eq:yprm}
\ee
where $y$ is a new parameter of order unity.  Note that the fractional
occupation number is exactly $1/2$ when $y=0$.
Inserting Eq.~(\ref{eq:yprm}) into Eq.~(\ref{eq:focdef}), we can now
take the limit as $L\rightarrow\infty$ to get
\bge
\lim_{L\rightarrow\infty}
 f_L\left(1+\frac{y}{2^L}\right)
 =
 \frac{1}{1-e^{-y}} -
 \frac{1}{y}.
\label{eq:finf}
\ee

Next, inserting Eq.~(\ref{eq:yprm}) into Eqs.~(\ref{eq:enta}) and
(\ref{eq:entb}), and taking the limit as $L\rightarrow\infty$, we find
the entropy excess,
\bge
\lim_{L\rightarrow\infty}
 \Delta S_L\left(1+\frac{y}{2^L}\right)
 =
 \ln\left(1-e^{-y}\right)-
 \frac{ye^{-y}}{1-e^{-y}}-
 \ln y + 1.
\label{eq:sinf}
\ee
Eqs.~(\ref{eq:finf}) and (\ref{eq:sinf}) constitute parametric algebraic
equations, with parameter $y$, yielding $\Delta S_L$ as a function of
the fractional occupation number $f_L$ as $L\rightarrow\infty$.  These
equations were used to produce the shaded curve in Fig.~\ref{fig:ent}.

Likewise, inserting Eq.~(\ref{eq:yprm}) into Eqs.~(\ref{eq:nois}), and
taking the limit as $L\rightarrow\infty$, we find the fractional noise,
\bge
\lim_{L\rightarrow\infty}
 \Delta {\cal F}_L\left(1+\frac{y}{2^L}\right)
 =
 \sqrt{\frac{y^2-2\cosh y + 2}{y^2-2y\sinh y + 2\cosh y -2}}
\label{eq:ninf}
\ee
Eqs.~(\ref{eq:finf}) and (\ref{eq:ninf}) constitute parametric algebraic
equations, with parameter $y$, yielding $\Delta {\cal F}_L$ as a
function of the fractional occupation number $f_L$ as
$L\rightarrow\infty$.  These equations were used to produce the shaded
curve in Fig.~\ref{fig:nois}.

Finally, inserting Eq.~(\ref{eq:yprm}) into Eqs.~(\ref{eq:gdefbrav}),
and taking the limit as $L\rightarrow\infty$, we find the $G_L$
factor for a Bravais lattice,
{\scriptsize
\bge
\lim_{L\rightarrow\infty}
 G_L\left(1+\frac{y}{2^L}\right)
 =
 \frac{
    2
   + \left(y^3+2y-8\right) e^y
   + \left(y^4-6y+12\right) e^{2y}
   + \left(y^4-y^3+6y-8\right) e^{3y}
   -2\left(y-1\right) e^{4y}}
      {
   \left[1 - (y^2+2) e^y + e^{2y}\right]^2}.
\label{eq:ginf}
\ee
}
Eqs.~(\ref{eq:finf}) and (\ref{eq:ginf}) constitute parametric algebraic
equations, with parameter $y$, yielding $G_L$ as a function of the
fractional occupation number $f_L$ as $L\rightarrow\infty$.  These
equations were used to produce the shaded curve in
Fig.~\ref{fig:gvsrho}.

\end{document}